Short Paper*

# The Perception of Filipinos on the Advent of Cryptocurrency and Non-Fungible Token (NFT) Games

Ryan D. Francisco
College of Engineering, Technological Institute of the Philippines – Quezon City, Philippines
rfrancisco.cpe@tip.edu.ph
(corresponding author)

Nelson C. Rodelas
College of Engineering, University of the East – Caloocan, Philippines
nelson.rodelas@ue.edu.ph

John Edison T. Ubaldo
University of the Philippines – Diliman, Philippines
jtubaldo@up.edu.ph

## Abstract

*Purpose* – This study aims to shed light on the rise of play-to-earn games in the Philippines alongside cryptocurrency. The lack of research and public understanding of its benefits and drawbacks prompted the researchers to investigate its market. As such, the study tried to look into the risks and benefits of crypto gaming if it would be regulated by the government, and how market volatility influences the churn rate of crypto games.

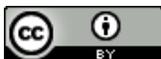

*Method* – The research used a descriptive study to determine the perception of people who are engaged in playing "Axie Infinity". The research also used non-probability sampling using the snowball method. The only tool used to collect data was Google Forms. Furthermore, the study used Statistical Package for Social Science (SPSS) to analyze the descriptive data.

*Result* – The results showed that most players spend their time playing Axie Infinity for about 1 to 4 hours a day. Predominantly, the return of investments for playing the game will take about 1 to 3 months. It also showed that these players agreed that there is a possible financial instability in a volatile market. With this, they have a high trust issue in terms of price manipulation, privacy and security, and its design and usability.

*Conclusion* – Understanding the cryptocurrency market requires comprehending the perspective of the people who are engaged in a play-to-earn game, and their concerns are critical for any government actions aimed at regulating self-employed income earners playing (Non-fungible Tokens) NFT games in the Philippines.

*Recommendations* – It is suggested that further research must be conducted to understand how self-employed Filipino income earners comprehend NFT gaming and cryptocurrency, impacting their level of interest and participation.

*Research Implications* – Taxing NFT games without sound regulatory frameworks may result in the disarray of this alternative income source. Therefore, the government must tread carefully to avoid a market collapse or other unfavorable outcomes.

*Keywords: Crypto-gaming, Churn rate, Cryptocurrency, Online Gaming, Axie Infinity, Perception*

## INTRODUCTION

When cryptocurrency was established in 2009, the world utilized it for virtual trading that resulting in the birth of more than 1,000 altcoins and crypto-tokens (Chuen, Guo, and Wang, 2017). Despite the promise that cryptocurrency showed, the Philippine market did not yet dive into this new intriguing venture. Unfamiliarity with the virtual trading of cryptocurrency and the nuances of this new market was the main obstacles that prevented Filipinos from tapping into this new resource.

However, recent innovations provided the country with a push towards cryptocurrency exploration through crypto-friendly technologies. In 2017, Bitcoin, the first decentralized cryptocurrency, became popular in the Philippines (England, 2021). Despite the breakthrough, the birth of cryptocurrency did not yet depict a significant impact on the socio-economic aspect of Filipino families as there were, at that point, limited resources and knowledge about the market. Unpredictability and volatility of the market were the main issues raised when looking at cryptocurrency market investments.

During the same year, competitive electronic sports (E-Sports) gaming made its way into the mainstream and became a trend, especially among young adults, due to the monetary implications involved. It also became more accessible to Filipinos due to the increased accessibility of serviceable mobile devices. Therefore, hopes and aspirations among Filipinos were on a high because of the opportunity that online gaming has now introduced.

The profile of end-game users affects the reception to online gaming. According to Penttinen, Halme, Malo, Saarinen, and Vilen (2017), there are two types of online gaming users: (1) primarily extrinsically-motivated and (2) primarily intrinsically motivated. Users who are intrinsically motivated enjoy completing tasks or missions without any sense of competition. With this, online gaming features such as in-game communications, easy-to-play mechanics, and usability are the priority of this kind of user. In contrast, extrinsically motivated users are goal-oriented. They are competitively motivated to consistently accomplish a goal or meet some externally imposed constraint (Hennessey, Moran, Altringer, and Amabile, 2015). This carries a heavy impact on how an online gamer performs within the context of online games. Regardless of the motivations, there is very little difference between the two user groups. The satisfaction that an online gamer can derive from playing is determined by their passion for completing tasks and missions.

The incorporation of cryptocurrency and online gaming was an unpopular livelihood source not until the COVID-19 pandemic hit in early 2020. With the world coming to a standstill, businesses were shut down resulting in a massive retrenchment and the lack of income among the middle class and below. In a third-world country like the Philippines, this was a dire situation that pressed Filipinos into exploring alternative sources of income. With the predicament came the rise of play-to-earn games such as Axie Infinity that paved the way for Filipinos to arrive at a new source of alternative income (Gill, 2021). Axie Infinity has become a popular investment avenue for Filipinos and also served as one of the leading sources of alternative income when the pandemic hit.

Despite the rise of this integrated cryptocurrency market to online gaming in the Philippines, it remains an enigmatic venture due to the lack of research and public awareness about its pros and cons. The current status quo still questions its full economic impact on Filipinos who are engaged in "crypto games".

This research aims to address the concern and hopes to shed light on this new venture that has penetrated the Filipino investment market. It will collate the experiences and perspectives of crypto gaming investors and players on the amount of time needed before they received their return of investment and the average hours of playing to be able to earn a decent income. Finally, the research will tackle possible risks and benefits of crypto gaming when regulated by the government from the perspective of investors and gamers alike, the trust level of users in cryptocurrency, and the impact of the volatility of the market on the churn rate of play-to-earn crypto games.

## LITERATURE REVIEW

### *Blockchain Cryptocurrency*

The situation that the world is currently facing led to the curiosity of researchers, developers, and investors to jump into blockchain-based cryptocurrency. According to Rehman, Salah, Damiani, and Svetinovic (2020), the menacing limited research study of cryptocurrency brings a heterogeneous trust issue to a different sector of technology providers and government. The study showed that there's a high-level trust issue in the aspect of price manipulation activities, volatility, transaction response time, ease of use, security, and privacy attacks.

Chuen, Guo, and Wang (2018) questioned the notion of some tax authorities where cryptocurrencies and tokens were classified as a commodity. Using the Cryptocurrency Index (CRIX), the study was able to understand and explore the risk and return characteristics of these altcoins that were classified as a commodity. Sentiment analysis was used to indicate the Sharpe ratio of the Cryptocurrency Index (CRIX). The result showed that the impact of blockchain cryptocurrency may extend beyond payments that are known as crypto-token offerings or token sales.

### *Risk and Opportunities in the Cryptocurrency Market*

With the promise of financial gain comes the risks in this new market. Canh, Thanh, and Thong (2019) provided a good overview of the main concerns that investors should consider before diving into the cryptocurrency market. The study found that structural breaks, like in many other investment markets, are also present in cryptocurrency, as well as volatility spillovers. This means that the cryptocurrency market has an intrinsic interconnectedness and lacks a diversified feature. Huynh (2019) agreed to the concern on spillovers and recommended the usage of different hedging instruments to address the issue of portfolio diversification.

Volatility, which is a characterization of any new market like cryptocurrency, can both be a good thing and a bad thing (Armour, 2021). Crypto value may rise and drop rapidly and thus the risk of gain and loss is too large to quantify and predict accurately. Despite the high level of volatility, the cryptocurrency market is still considered a good long-term investment that is advantageous because it is not restricted by borders and with an investor market accessible any time of the day (NDTV Business Desk, 2021).

### *Online Gaming*

According to Jang and Ryu (2009), the rise of competitive online gaming in the world "has become a major leisure activity." People who play online games such as MMORPGs (Massively Multiplayer Online Role-Playing Games) participate in multifaceted in-game interactions and some of which require a competitive skill to accomplish tasks and missions. With this, some online game players hire other players to join their organizations to perform strategic activities to deliver tasks together.

In addition to playing online games, some might question if playing can also be profitable. The study of Pentinnen et al. (2018) differs the user's performance based on their motivation. Testing the differences of primarily extrinsically motivated and primarily intrinsically motivated players. Primarily intrinsically-motivated players are those who just enjoy the game regardless of whether it is profitable or not. Meanwhile, primarily extrinsically-motivated players are those with a competitive mindset that will not falter a single day to complete certain tasks and missions to earn in the game.

*NFT Games and the Philippines*

Non-Fungible Token (NFT) games are a fairly recent development in the cryptocurrency market (Wang, Qin, Li, Wang, Qi, Chen, 2021), and its promise of income, coupled with economic difficulties in the Philippines, led to its rise in the Philippine gaming industry. As such, NFT games in the Philippines have effectively merged into the gaming industry (Manila Times, 2021). The Philippines even ranked first out of 20 countries in terms of NFT ownership according to a survey facilitated by Finder.com. Axie Infinity to become popular in Filipino households. The play-to-earn model that NFT games introduced led numerous Filipinos to explore this investment option and attracted players as well to engage in this venture. In 2021, Axie Infinity became a subject to various national news features due to the income generated from the game. Other NFT games such as Star Atlas, Splinterlands, and Illuvium also have enough engagement in the Philippines, but not on the level that Axie Infinity has penetrated the country. It is still considered by many as the most popular NFT game in the Philippines.

## METHODOLOGY

This descriptive study aimed to determine the perception of people who are engaged in a play-to-earn game, specifically the "Axie Infinity", that could help the development of crypto gaming in the Philippines. The research used non-probability sampling using the snowball method. This method of data collection gives the researchers one or more potential respondents. Data collection was solely facilitated using google forms, a popular online survey platform, to gather the data needed. The online survey form was distributed through different social media groups. People who have access to "Axie Infinity" served as the research locales of the study. The content-validated questionnaire was thoroughly counter-checked using Cronbach Alpha. The researcher's pilot-tested the research instrument to 50 respondents. These respondents were not part of the study. The items in the questionnaires revealed that were valid and reliable (Chronback's alpha ≥ 0.83). The questionnaire contains the respondents' age, average hours of playing Axie, minimum income in playing Axie, and the number of days playing Axie.

The respondents answered a total of 29 items that could be accomplished in about three to five minutes. Likert scale was employed to facilitate the analysis of the respondent's sentiments concerning the research agenda. Scale ranges of 1 (strongly disagree) to 5 (strongly agree) and 1 (Very Low) to 5 (Very High) were used separately in different parts of the questionnaire. For the questions that required a numeric answer, the respondents were asked to choose from a pre-set range prescribed by the researchers.

The questionnaire consisted of a total of 5 parts with two main sections: Demographic profile and Survey on Crypto-gaming. The demographic profile asked the respondents about the average per day playing hours. This section also sought to inquire about the length of time before the respondents were able to achieve their return-of-investment. However, the last question in the survey was specifically addressed to crypto investors of Axie Infinity only.

The second section consisted of four questions with pre-set possible responses. The choices under each question were derived from previous studies looking at cryptocurrency and online gaming. This section also gathered the personal perspective of the respondents regarding the possible risks and benefits of the crypto gaming venture if ever the government decides to regulate it. With the massive surge of Filipinos investing in the cryptocurrency market, the respondents were also asked to weigh the impacts of the volatility of cryptocurrency in crypto-gaming. Lastly, the questionnaire asked the respondents about their views regarding the reliability of cryptocurrency in the Philippines in terms of the following factors: Privacy and security, design and usability, governance and regulations, transparency, reputation, and parallel and the shadow economy.

Since the research used non-probability sampling using the snowball method, the number of respondents was solely dependent on the duration of the survey. The data were gathered from November 15 to November 18. With this, one hundred seventy-five responses were collected and these were all utilized in the study. The researchers used percentage, frequency count, and means in describing the data. To analyze the descriptive data, the Statistical Package for Social Science (SPSS) version 20 was used.

# RESULTS

Among the respondents, most of them are millennials and generation z (n = 84, 48%) who can play the game for about 1 to 4 hours a day (n = 139, 79%), see Figure 1. In addition, some lend their time for about 5 to 8 hours a day to complete certain missions and tasks in the game (n = 34, 19%).

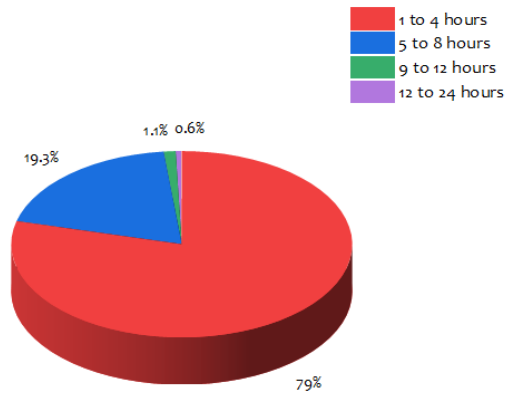

Figure 1. Number of hours spent playing Axie Infinity

Figure 1 shows that 79% of the respondents play between 1 to 4 hours per day while 19.3% play around 5 to 8hrs. The playing time corresponds to the required time that each player needs to be able to accomplish the daily tasks and meet the cut-off for them to maintain their average weekly income, which differs from each respondent. For most of them, playing for 1-4 hours a day is enough to maintain the continuous flow of income.

Figure 2 showed that predominantly, the return-of-investment will take about 1 to 3 months of playing the game (n = 69, 39%)

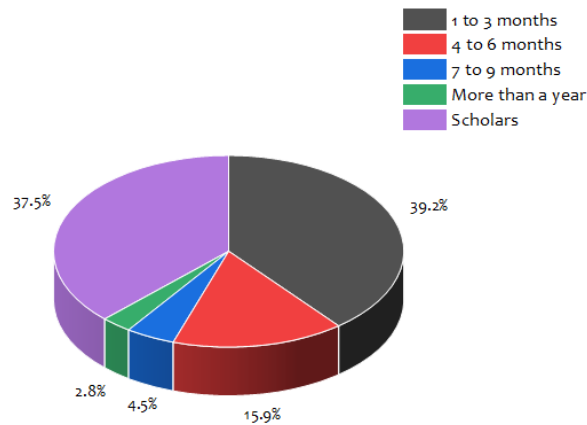

Figure 2. Number of months it took for the Return-Of-Investments

Table 1 shows that aside from the market's volatility, they also understand that there will be an increasing demand in the crypto assets resulting in financial instability (μ = 3.73, σ = 1.12).

Table 1. Possible risk of crypto gaming when regulated by the government

| Possible Risk | Mean (Min = 1, Max = 5) | Std. Deviation | Interpretation |
|---|---|---|---|
| Convenience Fee | 2.93 | 1.36 | Neutral |
| Financial Instability | 3.73 | 1.12 | Agree |
| Increase in volatility | 3.88 | 1.02 | Agree |
| Less accuracy in controlling the inflation rate in the currency | 3.51 | 1.01 | Agree |
| Congestions in transactions | 3.59 | 1.04 | Agree |
| Increase in churn rate | 3.48 | 1.08 | Agree |
| Low return-of-investment | 3.06 | 1.12 | Neutral |
| | μ = 3.45 | | Agree |

On the other hand, table 2 shows that most of these crypto gamers believe that there will be exponential economic growth once NFT games are regulated (μ = 3.55, σ = 1.00).

Table 2. Possible benefits of crypto gaming when regulated by the government

| Possible benefits | Mean (Min = 1, Max = 5) | Std. Deviation | Interpretation |
|---|---|---|---|
| Low latency (Time to make transactions) | 3.24 | 1.12 | Neutral |
| Reduce the level of pessimism of the speculators | 3.34 | 1.14 | Neutral |
| Increase in economic growth | 3.55 | 1.00 | Agree |
| Interest rate caps can be cancelled | 3.31 | 0.89 | Neutral |
| Data will be secured | 3.14 | 1.33 | Neutral |
| Controlled inflation rate in the currency | 3.37 | 1.02 | Neutral |
| Philippine Bank act as a lender of last resort and offer crypto loans | 3.23 | 1.15 | Neutral |
| | μ = 3.31 | | Neutral |

Contrarily, the players are most aware that one of the major impacts of volatility is a big investment gain (μ = 3.81, σ = 0.91). They also agree that higher volatility corresponds to a higher probability of a declining market.

Table 3. Impacts of volatility of the cryptocurrency market to the income generated from play-to-earn crypto games

| Impact of volatility | Mean (Min = 1, Max = 5) | Std. Deviation | Interpretation |
|---|---|---|---|
| Loss of investments | 3.19 | 1.18 | Neutral |
| Big investment gain | 3.81 | 0.91 | Agree |
| Higher volatility corresponds to a higher probability of a declining market | 3.59 | 0.96 | Agree |
| Churn rate will greatly increase | 3.37 | 0.96 | Neutral |
| | $\mu = 3.49$ | | Agree |

Table 4 shows that the unregulated income-generating crypto game and the lack of knowledge resulted in a high trust issue level in terms of privacy and security, design, and usability.

Table 4. Trust level of users in cryptocurrency and crypto gaming

| Trust issue | Mean (Min = 1, Max = 5) | Std. Deviation | Interpretation |
|---|---|---|---|
| Privacy and security | 3.72 | 0.99 | Agree |
| Design and usability | 3.92 | 0.86 | Agree |
| Governance and regulations | 3.28 | 1.05 | Neutral |
| Lack of transparency | 3.18 | 1.07 | Neutral |
| Reputation systems | 3.28 | 1.05 | Neutral |
| Parallel and shadow economy | 3.27 | 1.10 | Neutral |
| | $\mu = 3.47$ | | Agree |

## DISCUSSION

This study sought to analyze the experiences and perspectives of crypto gaming investors on the issues of return-of-investment, government regulation, risk and benefits, volatility, and trust level. To acquire empirical data for the analysis, the researchers gathered 176 responses from a mix of players and managers through an online survey. Although unintentionally, the majority of the respondents were male (75%) but this demographic was not a factor in the results of the study. The results showed that the respondents (composed mainly (96%) of individuals from the 10-19 and 20-29 yrs old age groups) were very much aware of the apparent risks involved in crypto gaming. It was also found that, for most of the respondents, the return of investment occurs within the next 1-3 months with an average playing time of 1-4 hours daily. The convenience of only having to play for a few hours while earning income that is almost as good as salaries in regular-paying jobs (Cruz, 2021) is what enticed a lot of Filipinos to explore opportunities within crypto gaming.

The respondents recognized that this crypto gaming venture comes with high rates of instability, volatility, and player congestion. They are generally aware of the financial instability and economic and financial volatilities as cryptocurrency/crypto gaming is not yet streamlined in the Philippines. The sentiments of the study's respondents reflect a similar reception among Axie players and investors on market volatility surveyed by One News, a Philippine news organization (Cruz, 2021). In the report, the respondents emphasized concern about market volatility before venturing into crypto gaming. Despite the dangers of volatility, the respondents were also expecting the other side of the curve with the possibility of experiencing massive investment gain.

Even with the uncertainties posed by the possibility of state regulation, the respondents believed that the market will experience growth considering its nature. This upside has become one of the major motivations of people that resulted in the recent surge in Filipino crypto gaming investments and the venture getting attention from national news agencies in 2021. Still, the data showed that players who engage in crypto gaming are mostly in it for short-term benefits (alternative income) and not for long-term investment. This is because, culturally, Filipinos are not into long-term investments. Income is the top priority among Filipinos and the lack of spare cash prevents them from committing to long-term investments (Tiongson, 2017). Overall, the topline knowledge of these players makes them undecided whether there will be a benefit in playing the game in the long run.

Outside of the unpredictability and volatile nature of cryptocurrency, the lack of regulation and lack of market knowledge resulted in high-level trust issues among the respondents of the research. With a dynamic market like a cryptocurrency that can be easily affected by outside intervention, state policies and regulations built upon personal/political agendas will be very detrimental to the development of an investment market (Reiter & Steensma, 2010). Philippine regulatory infrastructures have also been subject to scrutiny due to the inefficiencies of services, lack of staff, low funding, and corruption allegations (Wong 2009) which exponentially multiplies the volatilities already within the cryptocurrency market.

## CONCLUSIONS AND RECOMMENDATIONS

This paper has described the experiences and perspectives of crypto gaming investors on issues of market volatility, risk and benefits, and the possibility of government regulation. Crypto gamers are fully aware of the financial instability in the cryptocurrency market that may further worsen if the government decides to regulate it as crypto assets will soon be in great demand. Also, they believe that there is a possibility for exponential economic growth once crypto gaming is regulated. However, this study also showed that most players engaged in a "play-to-earn" game mainly for self-sustenance and not for the long-term development of the cryptocurrency in the Philippines. With this, their lack of substantial knowledge makes them unsure about investing in cryptocurrency for the long term.

The players are also aware that one of the major impacts of the market's volatility is the chance of big investment gain. Hence, their investment in playing the crypto game also depends on the volatility of the market. The higher the risk, the higher the returns when they hit their stride the right way. Furthermore, the unregulated income-generating crypto game and the lack of knowledge about the market resulted in a high trust issue level in the cryptocurrency market.

It is suggested that further research must be conducted to understand how self-employed Filipino income earners comprehend NFT gaming and cryptocurrency, impacting their level of interest and participation. It is also recommended to compare how government regulation affected other investment markets to juxtapose to the existing status quo within the cryptocurrency market.

## IMPLICATIONS

As our findings depicted, the cryptocurrency market is still very volatile, and adding taxation without concrete regulatory frameworks may result in the dissolution of an alternative source of income for Filipinos. Further studies need to be explored before applying regulation in this relatively new mode of income generation. Although regulation is a possibility and both managers and players welcome the idea, the government should tread carefully to avoid a market collapse or any undesired result.

## ACKNOWLEDGEMENT

The researchers are indebted to the respondents of the study. Also, we would like to personally thank Dr. Rex Bringula who provided us with his resources to assist us in developing this research. Lastly, we also thank the panelists who are always willing to help in the improvement of this paper.

**Author's Biography**

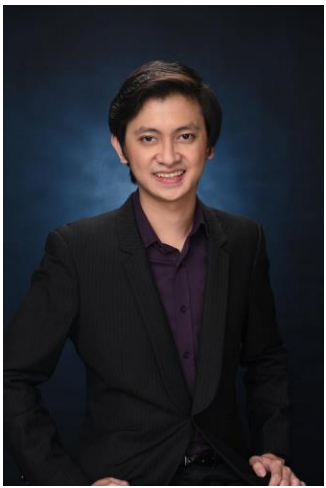

**Engr. Ryan Francisco, ES**

Ryan Francisco is a graduate of Bachelor of Science in Computer Engineering at the University of the East - Caloocan. During his undergraduate studies, he was a recipient of the Junior Level Science Scholarship of DOST in 2016. He is also one of the finalists of the Ten Outstanding Students of the Philippines for the National Capital Region. In addition, he won the GMA Excellence Award for the technology-based category in 2019. A faculty member at Technological Institute of the Philippines – Quezon City. His publication and research areas include data analytics, embedded system, and blockchain technology

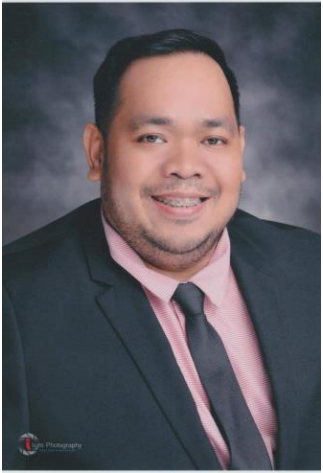

**Dr. Nelson Rodelas, PCpE**

Nelson is one of the proud faculty members of the College of Engineering under the Computer Engineering Department of the University of the East Caloocan at the same time a faculty member of the Graduate Studies in the ICT Department of the University of the East Manila. He is a graduate of Doctor of Information Technology at the University of the East, Manila, a Master of Science in Engineering major in Computer Engineering at the Polytechnic University of the Philippines, and a Bachelor of Science in Computer Engineering at the University of the East Caloocan. He is also a certified Board Passer in Licensure Examination for Teachers (LET) major in Mathematics. His publication and research areas include environmental research, waste management, data analytics, and deep learning.

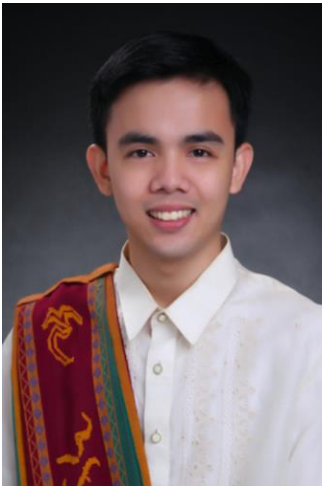

**John Edison Ubaldo**

Edison is a recipient of the Canada-ASEAN Scholarships and Educational Exchanges for Development (SEED) for the academic year 2020-2021 where he did a research exchange at the Centre d'études asiatiques - Université de Montréal. He has presented his papers focusing on agrarian reform in the Philippines and mining and development in multiple international conferences. Since January 2021, he has served as an Esports correspondent for GMA News Online covering the Mobile Legends Bang Bang Professional League Philippines.